%% file: twins.tex
\documentclass{article}
\usepackage[T1]{fontenc}
\usepackage{geometry}
\makeatletter

\providecommand{\LyX}{L\kern-.1667em\lower.25em\hbox{Y}\kern-.125emX\@}

\makeatother

\begin{document}

\title{On the Twin \emph{Non-}paradox}

\author{A. F. Kracklauer and P. T. Kracklauer}

\date{\date{}}

\maketitle
\begin{abstract}
It is shown that the ``twin paradox'' arises from comparing unlike entities,
namely perceived intervals with \emph{eigen}intervals. When this lacuna is closed,
it is seen that there is no twin paradox and that \emph{eigen}time can serve
as the independent variable for mechanics in Special Relativity.
\end{abstract}

\section{Introduction}

Informal remarks by editors of physics journals have it that by far the largest
number of submissions critical of contemporary physics attack Special Relativity.
Generally, these attacks call on the implicit absurdity of the relativity of
time and length spans; i.e., time-dilation and Fitzgerald contraction. Among
professional physicists, however, these concerns are most frequently written
off as the conservative stubbornness of amateurs unable to accommodate or comprehend
the force of `modern' mathematical reasoning.

Nevertheless, there are fully serious and absolutely rigorous arguments supporting
the amateur's heretical tendencies in this matter. Consider the twin paradox,
the core of which was actually first recognized by Einstein himself in the very
first article ever written on special relativity when he pointed out that Lorentz
transformations yield asymmetrical aging between various inertial frames.\cite{AE}
This feature was thereafter anthropomorphized by Longevin, who applied the principle
of asymmetric aging to the now fabled example involving twins, one of whom makes
a round trip while the other stays put.\cite{PL} As is very well known, conventional
analysis involving the Lorentz transformations seems to show that the the traveling
twin returns home to meet his sibling, now much older than himself. In so far,
however, as kinematically seen, both twins experienced a \emph{symmetric} relationship,
the time difference is paradoxical. Langevin resolved this paradox by calling
on the fact that the kinematical symmetry is broken by the \emph{dynamical}
fact that only the traveler experienced acceleration. Almost immediately, however,
von Laue observed that the effect was independent of the acceleration, as the
ageing effect could be extended simply by extending the length of the trip without
altering the accelerations involved; that is, the \emph{dynamical} aspects of
the trip must be irrelevant!\cite{ML} It can not be both ways, however---such
a conflict can not be ascribed to an unsophisticate's failure to appreciate
rigor. Ninety years, hundreds of books and thousands of articles later, despite
excursions prompted by all manner of considerations, this matter still stands
at exactly at the point von Laue left it.

Of course, special relativity has been verified by thousands of experiments;
its fundamental verity is unassailable. It is the purpose herein to propose
a potential resolution for this conflict therefore, which entails the minimum
ancillary modification to special relativity.

\section{Proper-length}

Previous analysis of the twin paradox has not carefully considered the issue
of the distance to the turn-around point (herein for brevity called the \emph{pylon})
of the traveling twin. This distance is not a vector on a Minkowski diagram,
but in fact the space-like separation of two entire whole world lines, namely
those of the terminus and of the pylon. The pylon, that is, its `place' in the
world, is not an event but a location. The turn-around itself is, of course,
an event in the usual meaning of that word for special relativity. For the traveling
twin, however, the turn-around event is a secondary matter as far as his navigational
needs are concerned. His primary concern is that he should travel to the correct
point in space, regardless of the time, before reversing course. How can he
do this? In the most natural way, he and his stay-at-home sibling chart a course
before the beginning of the trip; they select an object in the world, a star
say, and designate it as the turn-around pylon. From standard references they
know that this star is located in a particular direction at a determined distance
\( D. \) This distance is not the length of a Lorentz vector but the proper-length
of the displacement from the home location of the twins. With this in hand,
the traveling twin then determines the speed capabilites of his craft and calculates
the anticipated arrival time at the pylon. This information is interesting but
not vital, the traveler intends to proceed to the pylon regardless of the time
needed to arrive there.

The distance to the pylon star is not an apparent distance, the length of a
moving rod, for example, but the proper-length to the whole world line of the
selected star. Such a length is a scalar and is not to be tranformed by a Lorentz
transformation. The location of the world line of the pylon on a Minkowski diagram
depends on the axis to which it refers. That is, this world line with respect
to the stationary twin passes through the space coordinate at `\( D \)' on
the abscissa. Likewise, this world line must pass through the traveler's abscissa
also; but, because of the difference in the scale of the traveler's axis, this
same world line, although still parallel to the stay-at-home's world line, will
not be congruent to the pylon's world line referred to the stay-at-home's axis,
but is displaced by the scale factor. (It is this displacement that has been
overlooked in previous analysis and which distinguishes the approach taken herein.)
The consequence of this displacement is that, the intersection of the traveler's
world line with the world line of the pylon is found to be further out on the
traveler's world line than usually thought; i.e., the proper time taken to reach
the turn-around is seen to be greater than heretofore calculated. In fact, it
is equal to the proper time of the stay-at-home as he himself computes it for
the time taken by the traveler to reach the turn-around point. Thus, when the
whole trip is completed, both twins agree that they have experienced equal portions
of proper time since the start of the trip. Their \emph{reports} to each other
via light signals on the passage of time, in the usual way do not agree, however.
But they are such that the final totals at the end do agree.

These ideas are depicted graphically in Figure 1.

\input{Minchart.tex}

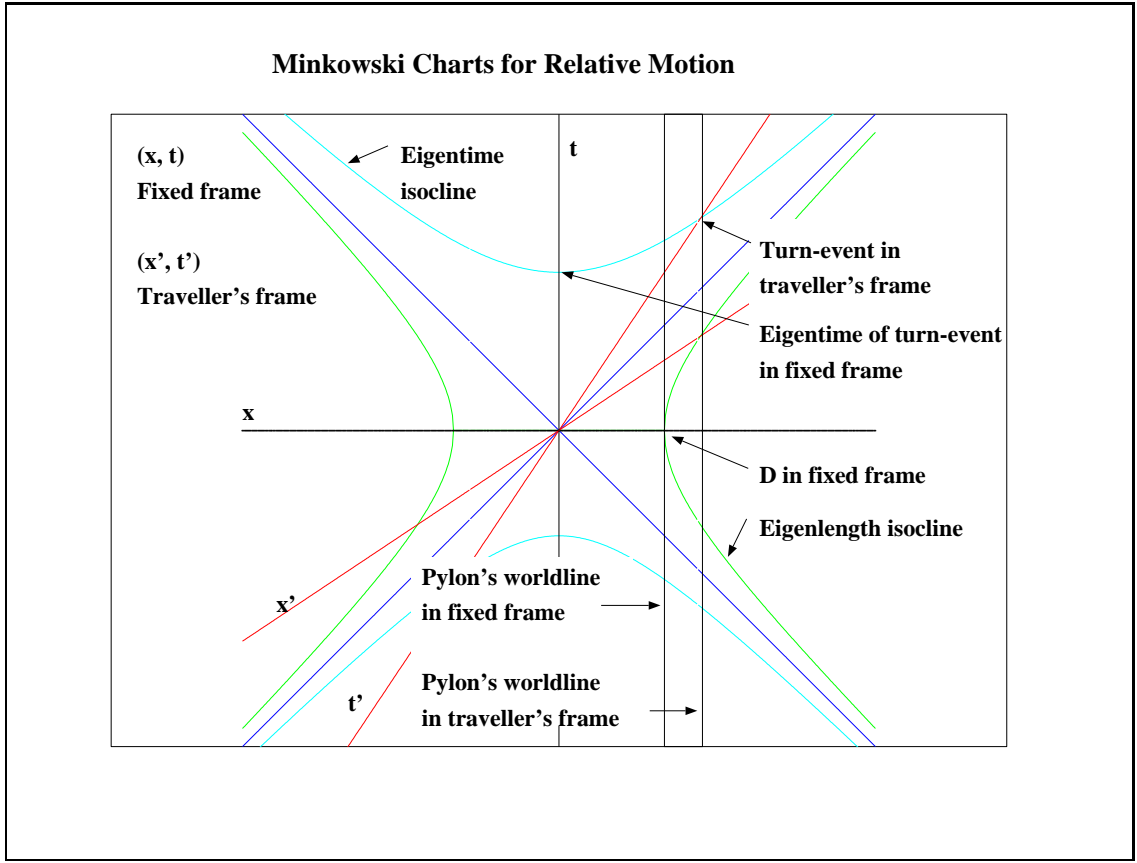
{\narrower\narrower\dessin{This figure is comprised of two Minkowski charts
superimposed on each other. The world line of the Pylon in the fixed frame passes
through the point `\( D \)' on the x-axis. The corresponding point on the x'-axis
is found by sliding up the \emph{eigen}length isocline to the intersection with
the x'-axis. The world line of the pylon passes through this point on the prime
chart. The intersection of the Pylon's world line with the t'-axis is the point
on the traveler's chart representing the `turn-around' event. The \emph{eigen}time
of the turn-around event in the fixed frame is found by sliding down that \emph{eigen}time
isocline which passes through the turn-around event to its intersection with
the t-axis. It is clear that this value is identical with the time assigned
by the fixed twin to the turn-around event as it may be projected horizontally
over to the intersection of the Pylon's world line in the fixed frame with the
time axis of the traveler. The paradox arises by using, incorrectly, that \emph{eigen}time
isocline which passes through the intersection of the traveler's and the pylon's
fixed frame worldlines. }{Figure}

}

\section{Experimental conflict}

All standard works on Special Relativity cite experiments attesting to the ``reality''
of time dilation and the effect yielding the twin aging discrepancy. How are
they to be understood in view of the above results? First, note that to date
no experiment meets the conditions leading to the twin-paradox. Certain experiments,
those involving muon decays, for example, are described by linear transformations
but are not round trips. ``Clocks-around-the-world'' experiments did involve
round trips, but not linear (acceleration free) motion. Further, note  that
time dilation is `real' in the sense that it actually occurs with respect to
signals. It is an effect attendant to `perspective' in space-time. Thus, all
physical effects resulting from the `appearance' (i.e., the way in which light
signals transmit information) will be modified by the the perspective. So any
test of time dilation which involves a report from or the interaction between
objects, will exhibit phenomena resulting from relative positions of emitter
and receiver; i.e., perspective.

Some experiments seem exempt from the effects of perspective. The two customary
examples are the muon decay curve in the atmosphere, and the transport of atomic
`clocks-around-the-world.' Here the situation is less clear. Each of these experiments,
however, is afflicted with features that allow contest.\cite{GG}

Muon decay, for example, largely seems to ignore possible cross-section dependence
on the velocity of the projectile and secondary production. The clocks-around-the-world
experiment has been strongly criticized for its data reduction techniques. Even
the existence of time delay effects for transported clocks has been questioned.\cite{EW}
Without access to the details of these experiments and their subsequent data
analysis, one is not in position to do deep critical analysis; nevertheless,
there is sufficient information in the literature to reasonably justify considering
conclusions drawn on their basis as disputable.

On the other hand, there are also experimental results completely in accord
with this result. An attempt by Phipps to observe the so called Ehrenfest effect---Fitzgerald
contraction of the circumference of a disk as a consequence of high tangential
velocity due to rotation---gave unambiguous null results, for example. \cite{TP}

\section{Conclusions}

This note is not derived from an effort to overthrow Special Relativity, rather
from an attempt to use it. Its fundamental point is that comparisons must be
made between like objects. Paradox results from the comparison of perceived
intervals as modified by space-time perspective with \emph{Eigen}intervals.
Filling this \emph{lacuna} in the understanding of Special Relativity enables
the resolution of a large number of conundrums similar to the ``twin paradox.''

The conclusions herein do not diminish the theory but actually extend its utility.
The arguments presented above obviously remain true when reduced to infinitesimals,
thereby enabling the piece-wise composition of an arbitrary (time-like) trajectory
in Minkowski space. They provide a substantiation of a resolution, proposed
by one of us in the past, of a deep problem in (special) relativistic mechanics
derived from the heretofore surmised lack of coordination among individual \emph{eigen}times
for interacting particles.\cite{AK} The considerations in this note constitute
a didactical elaboration of that argument in which it was observed that the
differential of arc length in Minkowski space is an invariant under Lorentz
transformation. That is, the differential of arc-length expressed in the instantaneous
rest frame along the orbit of the \( k \)-th particle at point \( p \), is
related to the differential of arc-length expressed in the instantaneous rest
frame of the differential of arc-length at any location \( p' \) on that or
any other arc \( j \) by a Lorentz transformation: \( \mathcal{L}(p,\: p',\: k,\: j) \):
\begin{equation}
\label{x-from}
dx_{k}|_{p}=\mathcal{L}(p,\: p',\: k,\: j)dx_{j}|_{p'}
\end{equation}

It follows, that the arc-length is an invariant as:
\begin{equation}
\label{e}
(dx_{k}|_{p}\cdot dx_{k}|_{p})^{1/2}=(dx_{j}|_{p'}\mathcal{L}^{T}(p,\: p',\: k,\: j)\cdot \mathcal{L}(p,\: p',\: k,\: j)dx_{j}|_{p})^{1/2}=(dx_{j}|_{p'}\cdot dx_{j}|_{p'})^{1/2}.
\end{equation}

This permits setting all such differential arc-lengths equal to a common expression:
\begin{equation}
\label{f}
c\, d\tau =(dx_{j}\cdot dx_{j})^{1/2},
\end{equation}

which can be rewritten as:
\begin{equation}
\label{g}
d\tau =\gamma ^{-1}_{j}dt_{j}\; \; \forall j,
\end{equation}
 where \( \gamma  \) has is customary meaning.

The arguments in this note give a more intuitively understandable rendition
of this fact by showing that whenever two world lines recross, \emph{eigen}intervals
starting from the previous recrossing, are equal, which is a restatement of
Eq. (\ref{g}). The utility of this fact for a theory of mechanics in Special
Relativity is exploited in Ref. (\cite{AK}).

\end{document}

%% file: Minchart.tex
%
 
 \long\def\Checksifdef#1#2#3{%
\expandafter\ifx\csname #1\endcsname\relax#2\else#3\fi}
\Checksifdef{Figdir}{\gdef\Figdir{}}{}
\def\dessin#1#2{
\begin{figure}[hbtp]
\begin{center}
\fbox{\begin{picture}(420.00,318.00)(20,0)
\includegraphics{\Figdir}
\end{picture}}
\end{center}
\caption{\label{#2}#1}
\end{figure}}